\documentclass[twocolumn,twoside,slac_two]{revtex4}
\usepackage{graphicx}
\usepackage{fancyhdr}
\usepackage{lscape}
\usepackage{multirow}
\usepackage{longtable}
\begin{document}

\newcommand{\gr}{$\gamma$-ray}
\newcommand{\grs}{$\gamma$-rays}
\newcommand{\vhe}{V\textsc{HE}}
\newcommand{\dg}{\ensuremath{^\circ}}
\newcommand{\nthu}{National Tsing Hua University}
\newcommand{\hess}{H.E.S.S.}

\title{Gamma-ray properties of globular clusters
and the ``fundamental planes''}

%

\author{P. Tam, A. Kong}
\affiliation{Institute of Astronomy and Department of Physics, National Tsing Hua University, Hsinchu, Taiwan}
\author{C. Y. Hui}
\affiliation{Department of Astronomy and Space Science, Chungnam National University, Daejeon, Republic of Korea}
\author{K S. Cheng}
\affiliation{Department of Physics, University of Hong Kong, Pokfulam Road, Hong Kong}

\begin{abstract}
We report on the discovery of gamma-ray emission from several globular
clusters (GCs), including Terzan 5, the second known gamma-ray GCs. By
now, more than a dozen GCs are known to emit gamma-rays of
energies above 100 MeV, thus enabling us to carry out the first
detailed correlation study with several cluster properties. We found
strong correlations between the observed gamma-ray
luminosities and four cluster parameters: stellar encounter rate,
metallicity [Fe/H], and energy densities of the soft photons
at the cluster locations. These ``fundamental planes'' of gamma-ray GCs
put an intimate relation of the observed gamma-rays to the underlying
millisecond pulsar population and have important implications on the origin of
the gamma-ray emission of GCs.
\end{abstract}

\maketitle

\thispagestyle{fancy}


\section{Introduction}
Radio and X-ray observations have revealed about 140 millisecond
pulsars (MSPs) in 26 globular clusters~\cite[GCs;][]{Freire_web}. However, the presence
of much stronger X-ray emitters can contaminate the X-ray observations
of MSPs. Because MSPs are the only known steady \gr~sources in GCs~\cite{lat_msp}, \gr~observations of GCs serve as an alternative channel in studying
the underlying MSP populations in GCs.

Using the Large Area Telescope (LAT), \grs~from 8 GCs~\cite{lat_8GCs} have been
discovered, including 47 Tucanae~\cite{lat_47Tuc} and Terzan 5~\cite{Kong_Terzan5}. 
\begin{figure*}
\includegraphics[width=75mm]{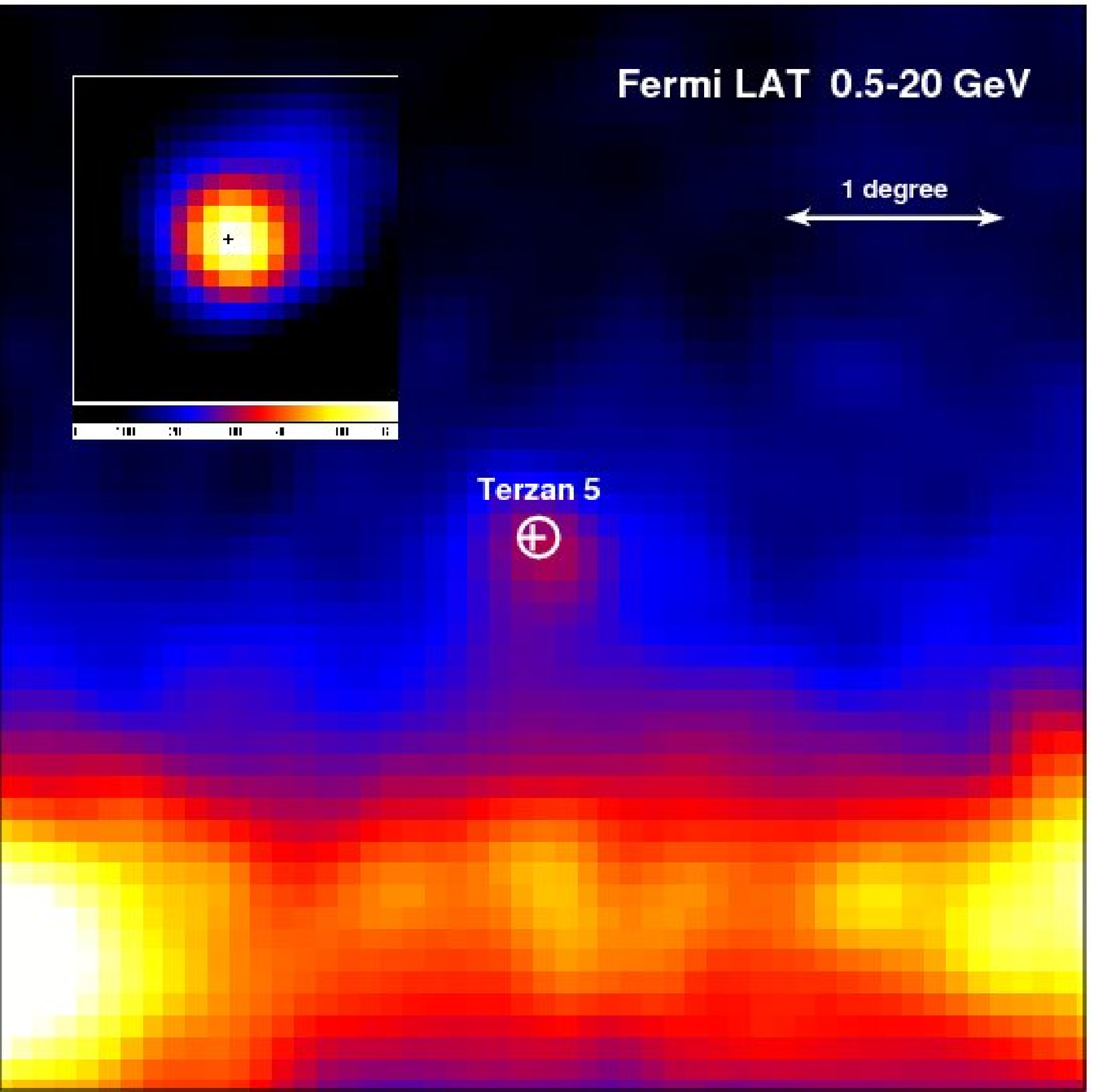}%
\includegraphics[width=75mm]{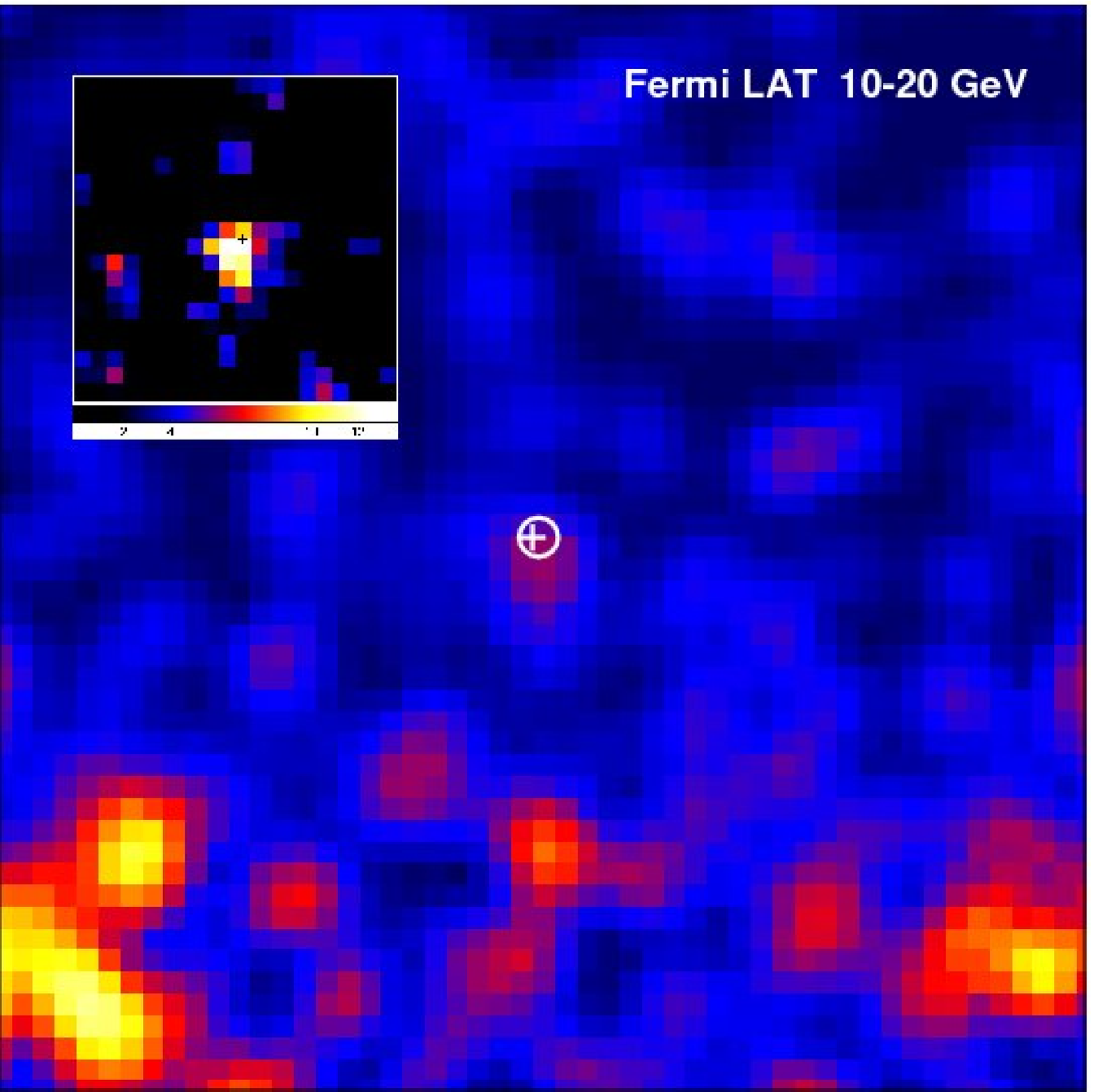}\\%
\caption{The count maps of the $5\dg\times5\dg$ region centered on Terzan 5. The insets show the test-statistic maps~\cite{Kong_Terzan5}.}
\end{figure*}

\section{Models of \grs~from globular clusters}
The radiation mechanism of \grs~is unclear. In the pulsar
magnetosphere model, e.g.~\cite{Venter08}, \grs~up to a few GeV come from the
MSPs through curvature radiation. On the other hand, inverse Compton
(IC) processes resulted from energetic particles up-scattering low-energy
photons, such as starlight and infrared light, may give rise to \grs~of
MeV to TeV energies, e.g.~\cite{Cheng_ic_10}. In either model, it is expected that the \gr~luminosity of a GC is proportional to the stellar encounter rate, a
measure of the number of MSPs in a GC.

\section{New \gr~globular clusters uncovered}
Terzan 5 contains the largest number of known MSPs among all GCs. It
was discovered as the second known \gr~emitting GC after 47~Tucanae~\cite{Kong_Terzan5} (see Figure~1). We note that 47~Tucanae was discovered in the bright source list~\cite{bsl_lat}, while the discovery of Terzan~5 in \grs~was announced~\cite{Kong_Terzan5} before the release of the first Fermi/LAT catalog~\cite{lat_1st_cat} and the report of the 8 GCs~\cite{lat_8GCs}.

Like 47~Tucanae, the \gr~spectrum of Terzan~5 also shows a cutoff at $\sim$3~GeV~\cite{lat_8GCs,Kong_Terzan5}.
After the discovery of other six \gr~emitting GCs~\cite{lat_8GCs}, we also identified
a group of GCs with high encounter rate. Using more than two years of data
taken from LAT, we found \gr~emission from the directions of Liller~1,
NGC~6624, and NGC~6752~\cite{Tam_gc_11}. The test-statistic maps of the regions around
these 3~GCs are shown in Figures~2 and 3. For M80, NGC~6139, and NGC~6541, the detection is marginal ($4-5\sigma$) when it was first reported~\cite{Tam_gc_11}.

For the cases where the \gr~emission is offset from the core (i.e. Liller 1
and NGC 6624), the \gr~spectra in the energy range of 200~MeV to 100~GeV are presented in Figure 4. The photons above $\sim$20~GeV are detected at significance levels of 3--4. Once the existence of these high-energy photons is established, it will be easier to be reconciled in the IC models than in the pulsar magnetosphere model. In the latter case, spectral cut-offs at several GeV are expected.

\begin{figure*}
\includegraphics[width=85mm]{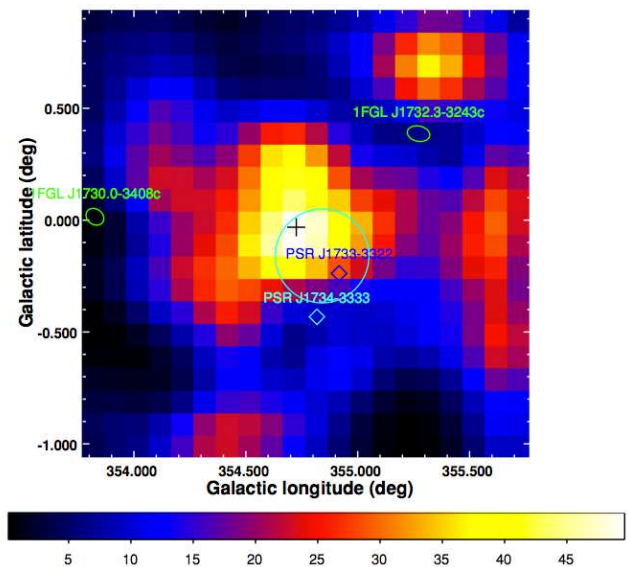}
\caption{The test-statistics map of Liller 1~\cite{Tam_gc_11}}
\end{figure*}

\begin{figure*}
\includegraphics[width=160mm]{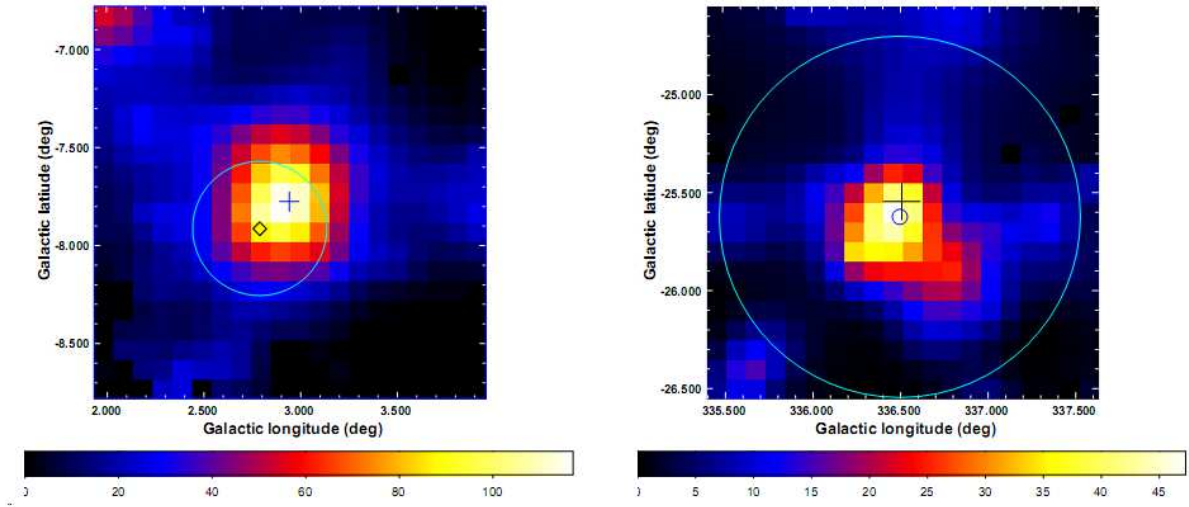}
\caption{The test-statistics maps of NGC~6624 (left) and NGC~6752 (right)~\cite{Tam_gc_11}}
\end{figure*}

\begin{figure*}
\includegraphics[width=80mm]{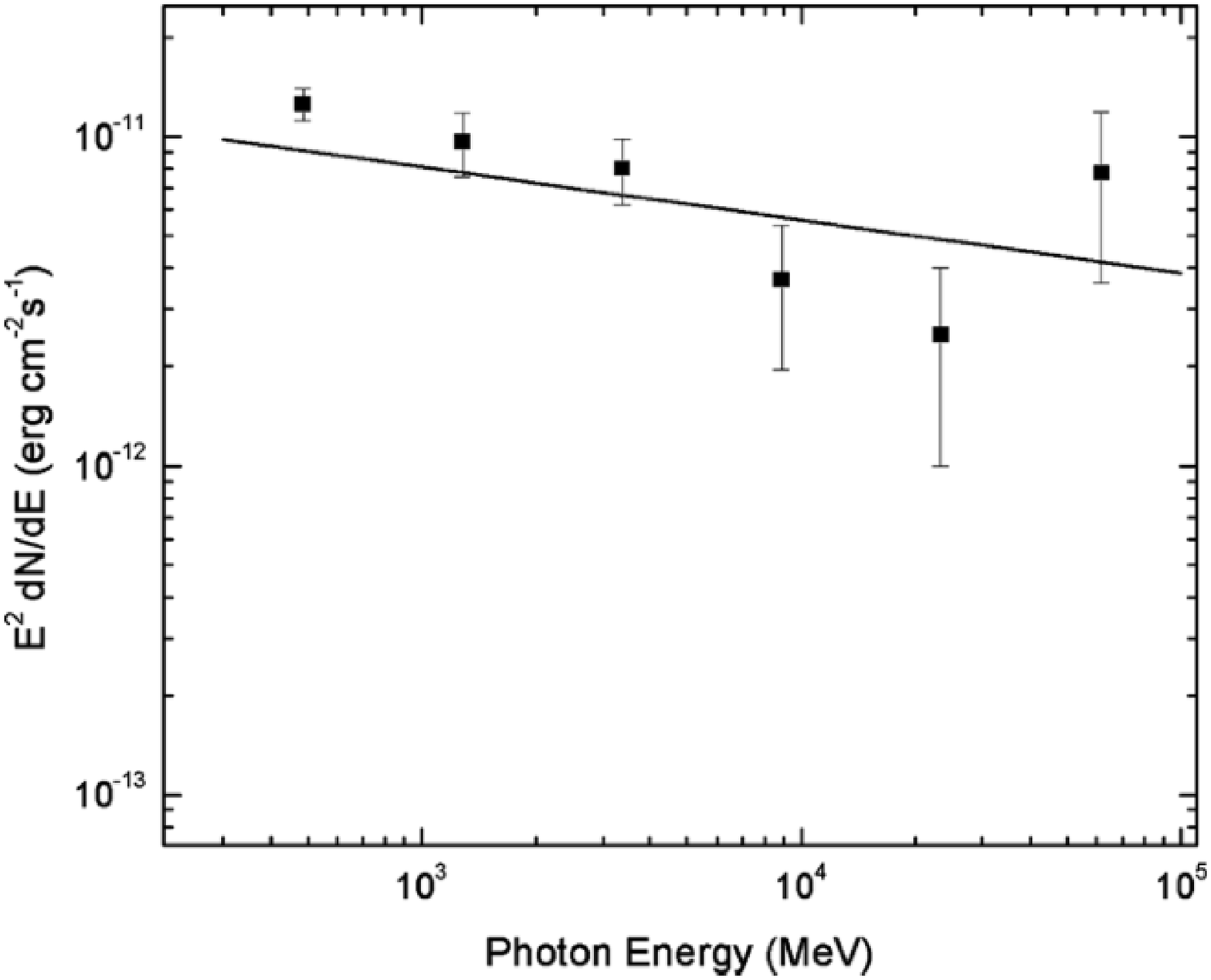}
\includegraphics[width=80mm]{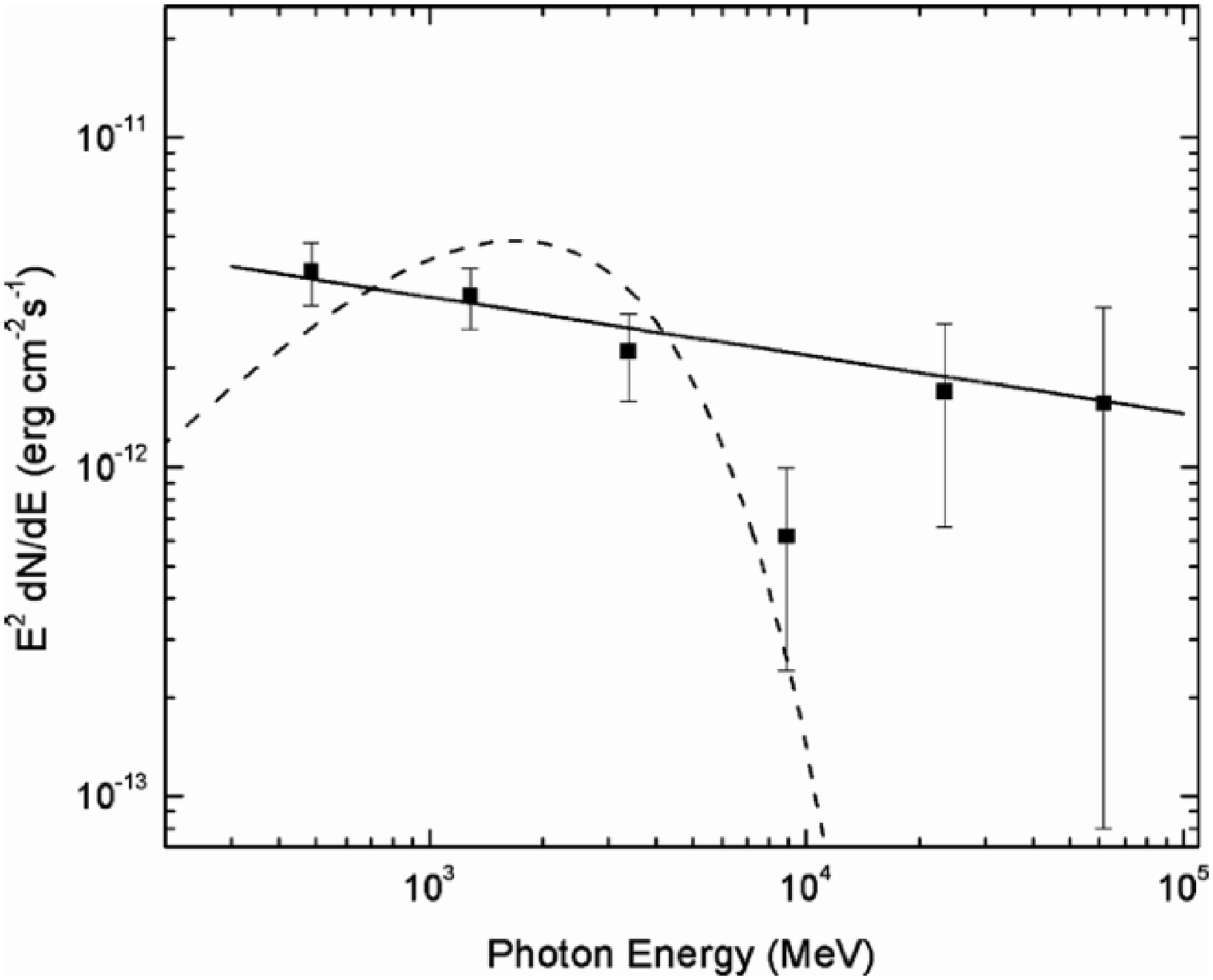}
\caption{Spectra of Liller 1 (left)
and NGC 6624 (right).
The solid and dashed lines
represent the best-fit
power law and power law
with exponential cutoff,
respectively~\cite{Tam_gc_11}.}
\end{figure*}

\begin{figure*}
\includegraphics[width=150mm]{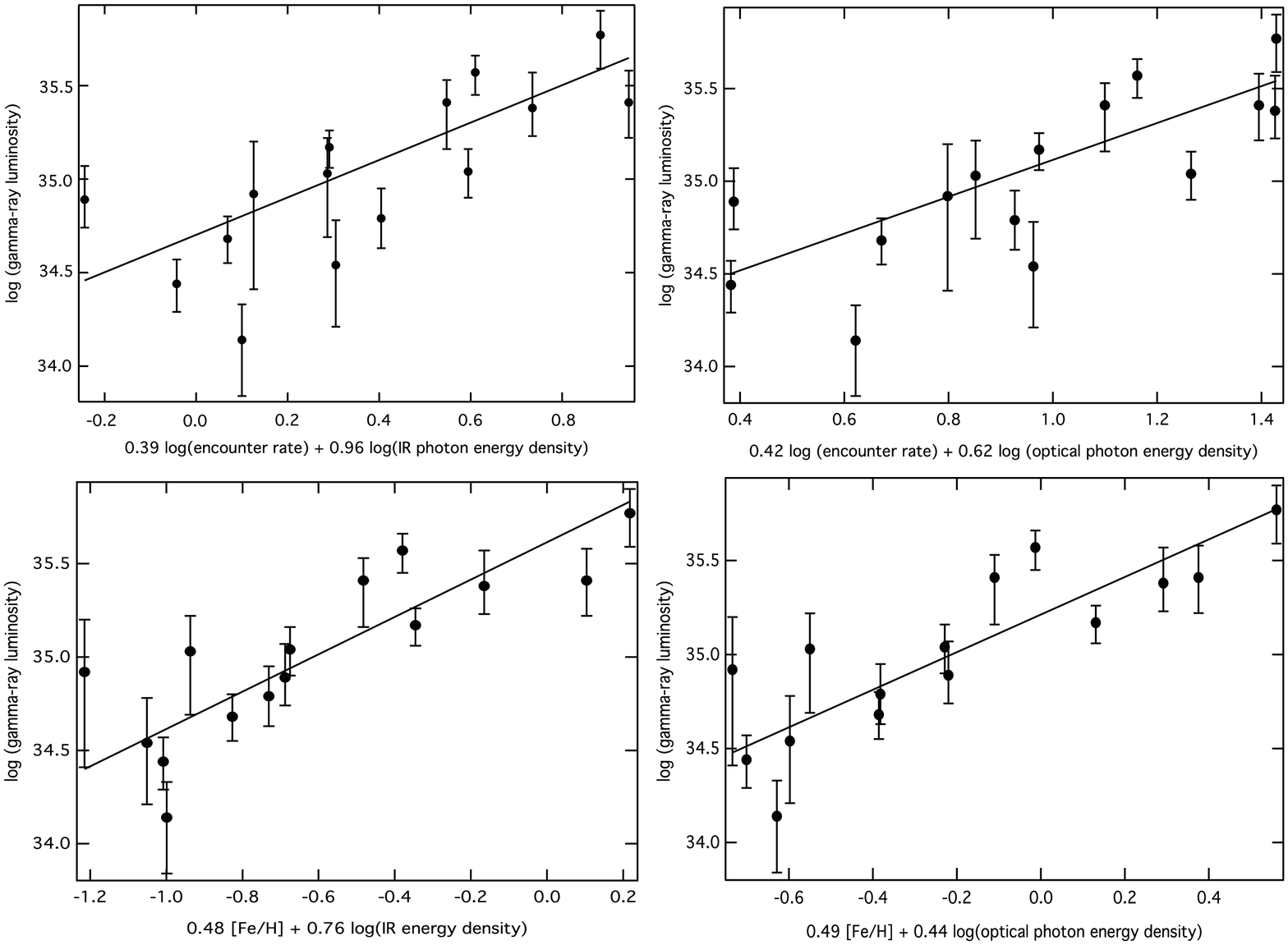}
\caption{The edge-on views of the fundamental plane relations of \gr~GCs. The straight
lines in the plots represent the projected best-fits~\cite{Hui11_correlation}.}
\end{figure*}

\section{The fundamental planes of \gr~globular clusters}

We have investigated the properties of the \gr~emitting globular
clusters~\cite{Hui11_correlation}. By correlating the observed \gr~luminosities with
various cluster properties, we probe the origin of the high energy
photons from these GCs. We found that the \gr~luminosity is
positively correlated with the encounter rate and the metalicity
[Fe/H] which places an intimate link between the \gr~emission
and the MSP population. We also found that the \gr~luminosity
increases with the energy densities of the soft photons at the
cluster location. When combining two parameters at the same
time, the correlation is even stronger. The edge-on fundamental
plane relations of \gr~GCs are depicted in Figure~5.

This finding strongly suggests that models that incorporate optical
or infrared photons should be taken into considerations in
explaining the \gr~emission from GCs, e.g. the IC models~\cite{Cheng_ic_10}.

\bigskip 
\begin{acknowledgments}
P. Tam acknowledges the support of the Formosa Program of Taiwan, NSC100-2923-M-007-001-MY3, and
the NSC grant, NSC100-2628-M-007-002-MY3. AK is supported by a Kenda Foundation Golden Jade Fellowship.

\end{acknowledgments}

\bigskip 

\end{document}